\documentstyle[epsfig,psfig]{elsart}
\begin{document}
\begin{frontmatter}
\title{Measurement of the cosmic ray hadron spectrum up to 30 TeV at
mountain altitude: the primary proton spectrum.} \centerline{ }
\centerline{The EAS-TOP Collaboration} \centerline{ }
\author[cnr,infn]{M.~Aglietta},
\author[infn]{B.~Alessandro},
\author[bologna]{P.~Antonioli},
\author[lngs,aquila]{F.~Arneodo},
\author[infn,uni]{L.~Bergamasco},
\author[infn,uni]{M.~Bertaina},
\author[uni]{C.~Castagnoli},
\author[cnr,infn]{A.~Castellina$^{*,}$},
\author[infn,uni]{A.~Chiavassa},
\author[infn,uni]{G.~Cini~Castagnoli},
\author[napoli]{B.~D'Ettorre~Piazzoli},
\author[napoli]{G.~Di~Sciascio},
\author[cnr,infn]{W.~Fulgione},
\author[infn,uni]{P.~Galeotti},
\author[cnr,infn]{P.~L.~Ghia},
\author[napoli]{M.~Iacovacci},
\author[cnr,infn]{G.~Mannocchi},
\author[cnr,infn]{C.~Morello},
\author[infn,uni]{G.~Navarra},
\author[infn]{L.~Riccati},
\author[infn,uni]{O.~Saavedra},
\author[cnr,infn]{G.~C.~Trinchero},
\author[infn,uni]{S.~Valchierotti},
\author[cnr,infn]{P.~Vallania},
\author[cnr,infn]{S.~Vernetto} and
\author[infn,uni]{C.~Vigorito}
\address[cnr]{Istituto di Fisica dello Spazio Interplanetario, CNR,Torino, Italy}
\address[infn]{Istituto Nazionale di Fisica Nucleare,Torino, Italy}
\address[bologna]{Istituto Nazionale di Fisica Nucleare,Bologna, Italy}
\address[lngs]{Laboratori Nazionali del Gran Sasso, INFN, Assergi (AQ), Italy}
\address[aquila]{Dipartimento di Fisica dell' Universit\`a, L' Aquila, Italy}
\address[uni]{Dipartimento di Fisica Generale dell' Universit\`a,Torino, Italy}
\address[napoli]{Dipartimento di Scienze Fisiche dell' Universit\`a and INFN, Napoli, Italy}

\date{\today}

\begin{abstract}                
The flux of cosmic ray hadrons at the atmospheric depth of 820
g/cm$^2$ has been measured by means of the EAS-TOP hadron
calorimeter (Campo Imperatore,
National Gran Sasso Laboratories, 2005 m a.s.l.).\\
The hadron spectrum is well described by a single power law : \\
$S_{h}(E_{h}) = (2.25 \pm 0.21 \pm 0.34^{sys}) \times 10^{-7}
(\frac{E_{h}}{1000})^{(-2.79 \pm 0.05)}$
m$^{-2}$s$^{-1}$sr$^{-1}$GeV$^{-1}$\\
over the energy range 30 GeV $\div$ 30 TeV. The procedure and the
accuracy of the measurement are discussed. \\
\vskip .8 cm \small (*) Corresponding author. \normalsize

The primary proton spectrum is derived from the data 
by using the CORSIKA/QGSJET code to compute the local hadron flux
as a function of the primary proton spectrum and to calculate and
subtract the heavy nuclei contribution (basing on direct
measurements). Over a wide energy range $E_{0}= 0.5 \div 50$ TeV its best 
fit is given by a single power law :\\
$S(E_{0}) = (9.8 \pm 1.1 \pm 1.6^{sys}) \times 10^{-5}
(\frac{E_{0}}{1000})^{(-2.80 \pm 0.06)}$
m$^{-2}$ s$^{-1}$ sr$^{-1}$ GeV$^{-1}$.\\
The validity of the CORSIKA/QGSJET code for such application has been
checked using the EAS-TOP and KASCADE experimental data by
reproducing the ratio of the measured hadron fluxes at the two
experimental depths (820 and 1030 g cm$^{-2}$ respectively) at
better than $10 \%$ in the considered energy range.
\end{abstract}

\end{frontmatter}

\section{Introduction}

The spectrum of hadrons detected at different atmospheric
altitudes retains significant information about the energy/nucleon
spectrum of primary cosmic rays, which is dominated by the
lightest component, i.e. the proton one. Its measurement has been
carried on in the past, both at sea level
\cite{dig,bro,bar,fic,ash,cow} and at mountain altitude
\cite{ame,can,ren,sio,ada}, using different experimental
techniques, like emulsion chambers, magnetic spectrometers and
calorimeters.\\
The knowledge of the primary proton spectrum is of main relevance
for the understanding of the cosmic rays acceleration mechanisms
and of the propagation processes in the Galaxy. Moreover, the
proton component is mainly responsible for the uncorrelated
particle production in the atmosphere: any uncertainties on the
proton spectrum reflect in an uncertainty in the calculation of the
secondary particle fluxes ($\pi$ and $K$) and thus, for example,
on the knowledge of the atmospheric muon and neutrino fluxes. A
precise knowledge of such spectra is of particular importance to
interpret the observational data from muon
and neutrino detectors deep underground \cite{cas}.\\
The measurement of the primary proton spectrum has been performed
by means of satellite and balloon borne experiments
\cite{boe,men,bel,san,web,alc,swo,buc,seo,rya,zat,iva,asa,apa} and
indirectly derived by using ground based detectors
\cite{ino,ame1,sak,aha}. In the region of tens of TeV, however,
direct measurements lack statistics and moreover their energy
determinations are not calorimetric and depend on the interaction
parameters and their fluctuations. The data inferred from
hadron measurements at ground level
can therefore provide significant new information.\\
On the other side, the derivation of the information on the
primary cosmic ray spectrum from hadron measurements, as well as
the comparison of the results from different experiments, relies
on the use of simulation tools describing the interaction and
propagation of primary cosmic rays in the atmosphere. The response
of such hadron interaction models has therefore to be verified,
especially considering that the recorded hadrons are the results
of large fluctuations with respect to the
average behavior. \\
The EAS-TOP Extensive Air Shower array was located at Campo
Imperatore, 2005 m a.s.l., above the underground Gran Sasso
Laboratories, with the aim of studying the cosmic ray spectrum in
the energy range $10^{13} \div 10^{16}$ eV through the
detection of the different air shower components.\\
In this paper, we present and discuss the results obtained in the
study of the uncorrelated hadrons by means of the calorimeter of
EAS-TOP,
namely: \\
a) the measurement of the hadron flux in the energy range 30 GeV
$\div$ 30
TeV;\\
b) the derivation of the primary proton energy spectrum in the
range 0.5 $\div$ 50 TeV; \\
c) the check of the propagation and interaction code
(CORSIKA/QGSJET) used for the interpretation of the data.

\section{\bf The detector and the hadron trigger}

\begin{figure}[h]
 \begin{center}
  \mbox{\psfig{figure=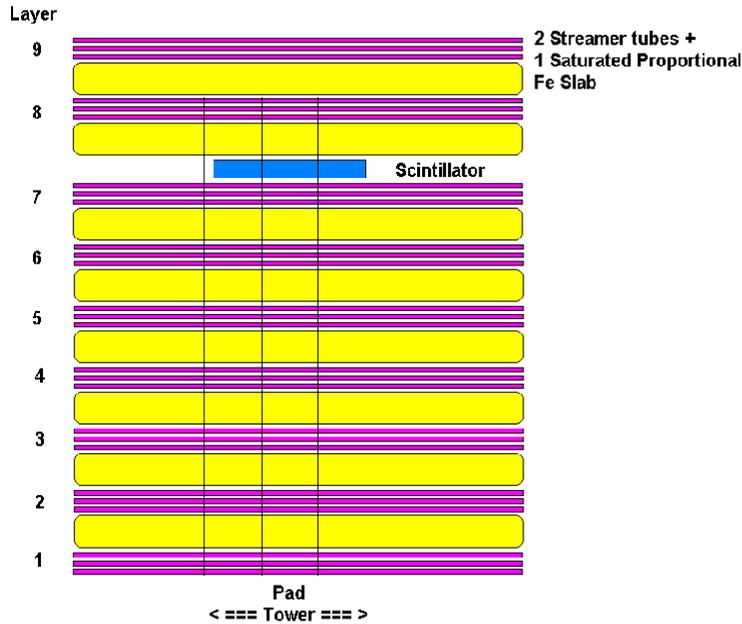,width=10.0cm,height=8.5cm}}
 \end{center}
\caption{\em {Schematic view showing the plane numbering and one of the 
defined ``towers'' inside the EAS-TOP calorimeter.}}
\label{fi:mhdsk}
\end{figure}

The Muon and Hadron Detector of EAS-TOP is a 144 m$^2$ calorimeter
\cite{nim} made of nine layers, each composed by a 13 cm iron slab
absorber (except for the uppermost plane which is unshielded), and
three planes of limited streamer tubes, for a total depth of 818.5
g cm$^{-2}$.\\ Two of the streamer tube layers (with 100 $\mu$m
wire diameter at 4650 V voltage) act as tracking devices,
and are read by a two-dimensional system based on the anode wires
and external orthogonal pick up strips. The third one, which data
are used for the present analysis, operates in saturated
proportional mode (the wire diameter being 50 $\mu$m, and the HV
at 2900 V) for hadron calorimetry and EAS core studies. The
signal charges are collected by a matrix of 840 (40$\times$38)
cm$^2$ pads placed above the tubes; the pad signals are
transferred to charge integrating ADCs with 15 bit dynamic range.
The pad read-out is converted to the equivalent number of vertical
particles by means of periodical calibration runs based on single
muon triggers (pressure and temperature dependence of the induced
charge being corrected for). \\ 
Different sets of scintillators
are placed in the apparatus for different aims; in particular, of
the six ones lodged below the second absorber layer, three are
used for hadronic trigger purposes. Each scintillator, of
dimensions (80$\times$80) cm$^2$, is centered on a pad, viewed by
two identical photomultipliers operating in coincidence and
discriminated at the level of 30 m.i.p., corresponding to the
energy loss of a 30 GeV proton incident on the calorimeter. They
provide the ``local hadron trigger'', which generates the read-out
of the whole detector.\\ 
For each scintillator a "tower" is defined, as the stack of 
$3 \times 3$ pads of the 8 internal layers centered on the scintillator
itself.
The detector and its operation are fully described in ref.
\cite{nim}; a scheme of a "tower" is shown in Fig.\ref{fi:mhdsk}.

\section{Hadron selection, acceptance and energy calibration}

\begin{figure}[bph]
\begin{center}
  \mbox{\epsfig{figure=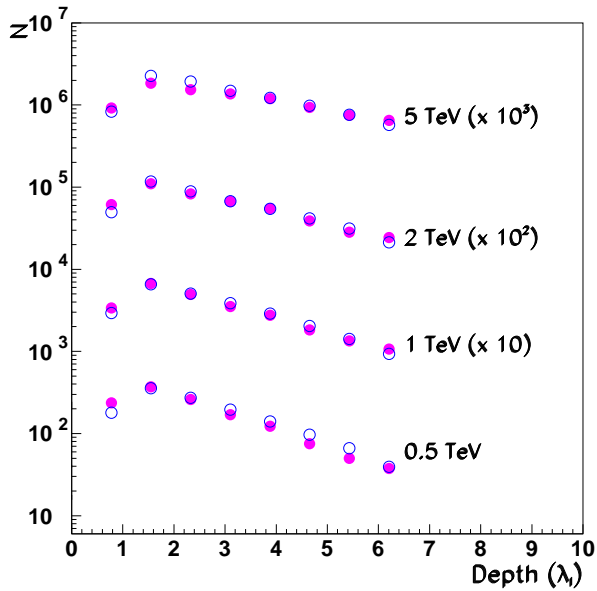,width=8.5cm,height=8.5cm}}
\end{center}
\caption{\em {Mean transition curves for hadrons in the
calorimeter. Full circles: experimental data; empty circles:
simulation.}} \label{fi:longi}
\end{figure}

An event, recorded following the ``local hadron trigger'', is
accepted as a hadron if: a) the cascade crosses at least three
consecutive internal layers of the calorimeter, including the one
positioned immediately below the triggering scintillator, and b)
the maximum energy release is recorded on the central pad of each
plane of the corresponding "tower". This allows the selection of
hadrons with energy above 30 GeV, and the definition of the
angular acceptance.\\
The check of hadron selection, the detection efficiency, effective
area, angular acceptance, and energy calibration have been
obtained by means of simulations
of the detector response based on the GEANT code \cite{mcgea}
(with FLUKA option for hadronic cascades), including the full
description of the apparatus. Protons at fixed primary energy and
zenith angle have been generated and analyzed with the same
procedure as the experimental data. \\
Particular care has been put in the modelling of the chamber
behavior in the saturated proportional mode; the saturation in the
collected charge has been studied in detail and included in the
simulation, as fully discussed in \cite{nim}. The modelling of the
chamber response to large particle densities has been checked at the 
50 GeV $e^+$ beam at CERN-PS, using a detector built by
chambers with the same characteristics, read out and filling gas
mixture as the ones operating on site, but with length reduced
to 3 m. Lead was used as absorber in front of the chambers in
order to reach maximum particle density \cite{nim}. The chamber
response was tested and found to agree with the model inside
2$\%$ up to particle densities $\rho_{ch} \simeq$ 300
particles/cm$^2$, corresponding to a 50 GeV electromagnetic shower
after 4 cm of lead absorber. For iron absorber and the calorimeter
geometry such particle density corresponds to hadrons with
energy $E_{h} \simeq$ 650 GeV.\\
Such response, introduced in the quoted simulation, provides
transition curves that can be directly compared with the
experimental data. As shown in Fig.\ref{fi:longi}, the difference
between the two curves 
is always less than 10 $\%$ even at shower
maximum, where the particle density is the highest, thus showing
that the chamber behavior and saturation are well described at
least up to 5 TeV (i.e. at particles densities at which
the chamber response could not be directly tested).

\begin{figure}[h]
 \begin{center}
  \mbox{\epsfig{figure=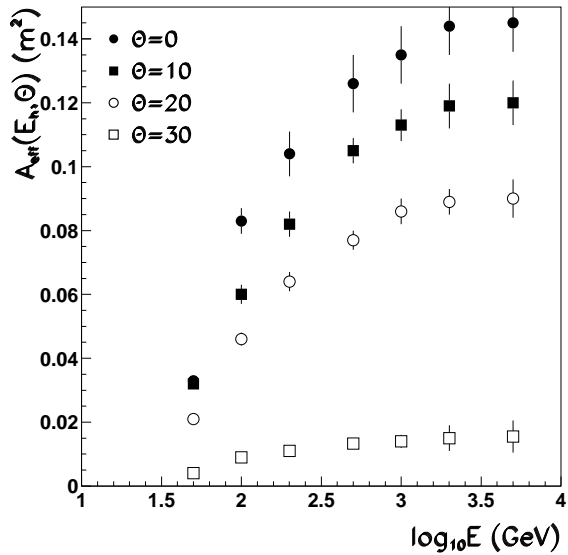,width=8.5cm,height=8.5cm}}
 \end{center}
\caption{\em {Effective area $A_{eff}(E_{h},\theta)$ for each
``tower'' vs $log_{10}E_{h}$ for 4 different hadron incidence
angles.}} \label{fi:acce}
\end{figure}

The verification of the hadron selection procedure has been performed
by comparing the shapes of the longitudinal developments for
individual events with the expectations from the simulated ones
(the agreement on the average transition curves having been shown in
Fig.\ref{fi:longi}). For layers 1-7 (i.e. the ones shielded by more than two
iron slabs, see Fig.\ref{fi:mhdsk}) the experimental
and simulated $N_{l}/N_{tot}$ distributions (i.e. the ratio between the 
equivalent particle number recorded in each layer and the total
one in the tower) are in agreement inside the statistical errors
($\simeq 10\%$). For layer 8 (shielded by a single iron slab), the
contamination from the accompanying shower adds an excess of 15$\%$ of
$N_{tot}$ in 16$\%$ of the events, independent on $N_{tot}$. The effect
does not alter the hadron selection and the spectrum measurement 
beyond the systematic effects discussed in the following.\\
For the described triggering conditions, the effective area
$A_{eff}(E_{h},\theta)$ was determined using the same simulation code and
taking into account the inefficiency of the trigger scintillators
due to the 30 m.i.p. threshold. Such area includes the detection
efficiency, which, concerning energy, rises above 65 $\%$ at 
$E_{h} \simeq$ 130 GeV for vertical incidence inside the geometrical area 
of the central pad. As regards zenith angle, the efficiency at 30$^{\circ}$ 
is about 10 $\%$ of the vertical one. 
The selection condition therefore introduces a cut in the
angular acceptance such that 90$\%$ of the events are found inside
22$^{\circ}$ from the vertical direction.\\
The effective area of each ``tower'' is shown in Fig.\ref{fi:acce}
for 4 different zenith angles. 

\begin{figure}[bph]
 \begin{center}
  \mbox{\epsfig{figure=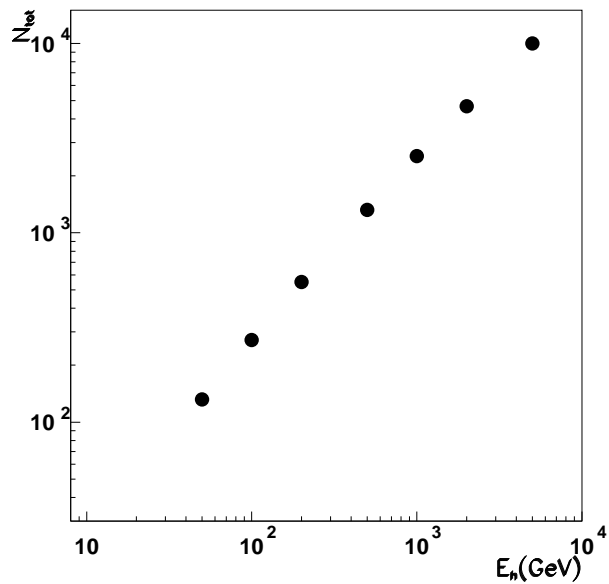,width=8.5cm,height=8.5cm}}
 \end{center}
\caption{\em {Total number of induced equivalent particles versus
primary energy.}} \label{fi:neconv}
\end{figure}

The hadron energy is inferred from the total charge induced on
the 8 shielded layers of the defined "tower" (more than 95 $\%$ of
the shower particles at all energies are contained inside a 20 cm
radius from the hadron position).\\
The conversion curve from the total number of particles induced in
the "tower" ($N_{tot}$) to the primary hadron energy is shown in
Fig.\ref{fi:neconv}. The energy resolution is $\sigma(E_{h})/E_{h}
= 15 \%$ at 1 TeV, worsening to 25$\%$ at 5 TeV due to leakage
losses and to 30$\%$ at 30 GeV due to sampling losses.
The dependence of the total number of particles on the hadron
zenith angle is less than 3$\%$ up to 30$^{\circ}$; the difference in 
the conversion curve between
protons and pions impinging on the calorimeter is less than 2$\%$.

The possibility that the triggering and selection procedure
includes more hadrons has been studied by means of a simulation of cascades 
in the atmosphere through CORSIKA/QGSJET. 
It results that such hadron pile-up effect,
even at the highest energies ($E_{h}>$3 TeV), does not alter the average energy
determination of more than 6$\%$. As a test,
to evaluate possible contaminations from the accompanying shower particles,
the particle-energy conversion curve has also been obtained using
the total charge induced on the five innermost planes only. No statistically
significant difference was found in the hadron fluxes obtained in
the two cases over the considered energy range.

\section{\bf The hadron flux}
\label{sec:hadflu}

About one million triggers were recorded in T=14760 hours of
effective live time used in the present analysis; 40832 of them
survived the selection criteria and were classified as hadrons. \\

The measured number of events in each energy bin for the flux
$S(E,\theta)$ is :
\begin{eqnarray}
n_{ev}^{meas}(E_{h} \div E_{h}+\Delta E_{h}) = \int_{0}^{\Omega}
\int_{E_{h}}^{E_{h} + \Delta E_{h}} S(E,\theta)T A_{eff}(E,\theta)
d\Omega dE
\end{eqnarray}
The hadron flux at zenithal angle $\theta$ can be approximated as:
\begin{equation}
S(E,\theta) = S(E) exp \big[-\frac{x(\theta)-x}{\Lambda(E)} \big]
\label{eq:lamflu}
\end{equation}
where S(E) is the flux in vertical direction and $x(\theta)$ is the
atmospheric depth along $\theta$.
The attenuation length $\Lambda(E)$ has been derived using the 
CORSIKA \cite{corsi} code
to simulate the interactions and propagation of primary protons in
air. In fact the hadron flux in the atmosphere includes both 
residual primaries and secondaries; at the EAS-TOP atmospheric depth, 
their ratio rises from $\simeq$0.7 at 500 GeV to $\simeq$1.4 at 5 TeV.
Therefore the obtained values of $\Lambda(E)$
represent the full evolution of such mixture: $\Lambda(E) \simeq$
114 g/cm$^2$ for QGSJET \cite{qgs}, and $\simeq$ 131 g/cm$^2$ for 
HDPM \cite{hdpm}, for the EAS-TOP altitude and range of zenith angles.
\\Assuming a power law spectrum ($\gamma$ = 2.7) inside each
energy bin, the mean value $\overline{E}_{h}$ is obtained, the
corresponding flux being
$S(\overline{E}_{h})=S(E)(\frac{\overline{E}_{h}}{E})^{-\gamma}$
(a change of $\Delta \gamma=0.1$ in the spectral slope does not
produce any appreciable difference in the resulting flux).\\
The vertical flux is thus :
\begin{eqnarray}
S(\overline{E}_{h}) = \frac{n_{ev}^{meas}(E_{h} \div E_{h}+\Delta
E_{h})} {2 \pi T \overline{E}_{h}^{\gamma}\int \int E^{-\gamma}
exp \big[-\frac{x(\theta)-x}{\Lambda(E)} \big] A_{eff}(E,\theta)
sin\theta d\theta dE} \label{eq:eqflu}
\end{eqnarray}
The recorded number of events and the experimental hadron fluxes
at the atmospheric depth of 820 g/cm$^2$ are listed in
Tab.\ref{ta:tabflu} with the corresponding statistical
uncertainties.

\small
\begin{table}[!tbp]
\centering
\begin{tabular}{|c|c|c|c|c|c|} \hline
{\bf Mean Energy } & {$\bf{E_{0}}$} & {$\bf{E_{1}}$} & {\bf Hadron} & {$\bf{S_{had}}$ } & {$\bf{\sigma(S_{had})}$} \\
{\bf  (GeV) } & {\bf (GeV) } & {\bf  (GeV)} & {\bf numbers} &
{$\bf{( m^{2} \ s \ sr \ GeV)^{-1}}$} & {$\bf{(m^{2} \ s \ sr \
GeV)^{-1}}$}  \\ \hline
    41 &    32 &    56 & 10222 & $.12 \ 10^{-2}$ & $.12 \ 10^{-4}$ \\
    73 &    56 &   100 & 12875 & $.27 \ 10^{-3}$ & $.24 \ 10^{-5}$ \\
   129 &   100 &   178 &  9506 & $.60 \ 10^{-4}$ & $.63 \ 10^{-6}$ \\
   229 &   178 &   316 &  4930 & $.14 \ 10^{-4}$ & $.21 \ 10^{-6}$ \\
   408 &   316 &   562 &  2174 & $.29 \ 10^{-5}$ & $.66 \ 10^{-7}$ \\
   726 &   562 &  1000 &   802 & $.47 \ 10^{-6}$ & $.18 \ 10^{-7}$ \\
  1290 &  1000 &  1778 &   299 & $.92 \ 10^{-7}$ & $.56 \ 10^{-8}$ \\
  2295 &  1778 &  3162 &   119 & $.17 \ 10^{-7}$ & $.16 \ 10^{-8}$ \\
  4081 &  3162 &  5623 &    44 & $.26 \ 10^{-8}$ & $.46 \ 10^{-9}$ \\
  7257 &  5623 & 10000 &    23 & $.84 \ 10^{-9}$ & $.18 \ 10^{-9}$ \\
 12904 & 10000 & 17783 &    12 & $.14 \ 10^{-9}$ & $.49 \ 10^{-10}$ \\
 22945 & 17783 & 31623 &     3 & $.55 \ 10^{-10}$ & $.39 \ 10^{-10}$ \\ \hline
\end {tabular}
\caption{\em{The measured hadron flux at 820 g/cm$^2$. The given
uncertainties are the statistical ones.}} \label{ta:tabflu}
\end{table}
\normalsize

The following sources of systematic uncertainties have to be
considered:

\begin{itemize}
\item[1. ] The uncertainty in the evaluation of the effective area,
$\frac{\delta A_{eff}}{A_{eff}} \simeq 12 \%$ at all energies.
\item[2. ] The uncertainty in the hadron angular distribution,
which reflects in the evaluation of the attenuation length
$\Lambda(E)$. A comparison between two different models (QGSJET
and HDPM) in the CORSIKA frame shows that
the differences in $\Lambda(E)$ reflect in a flux uncertainty
$\frac{\delta S}{S} \simeq 5 \%$.
\item[3. ] The uncertainty in the energy assignment to each single hadron, 
for a spectral slope $\gamma \simeq$ 2.7, results in a flux
uncertainty $\frac{\delta S}{S} \simeq 7 \%$. This value reaches
10$\%$ at the highest energies, as shown by the comparison of the
measured and simulated longitudinal shower profiles.
\item[4. ] An uncertainty $\frac{\delta S}{S} \simeq 15 \%$ on the flux,
due to the different behavior and efficiency of the triggering
scintillators and to the different calibrations and stability of
the corresponding "towers".
\end{itemize}

\begin{figure}[tbp]
 \begin{center}
   \mbox{\epsfig{figure=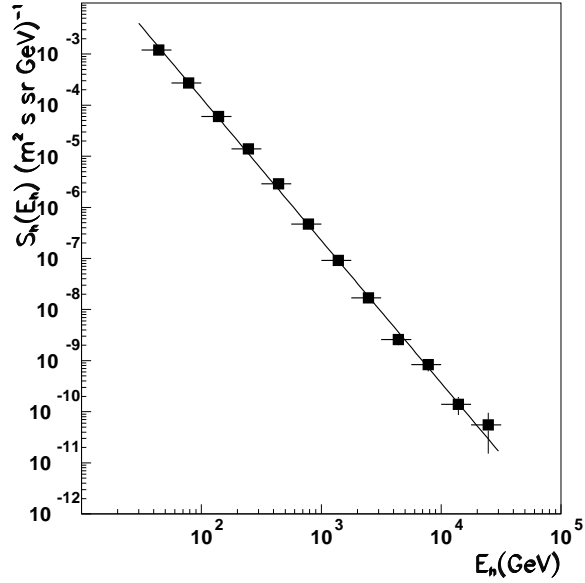,width=8.5cm,height=8.5cm}}
 \end{center}
\caption{\em {The hadron flux at 820 g/cm$^2$. The best fit
(\ref{eq:flufit}) is shown superimposed to the data.}}
\label{fi:hadflu}
\end{figure}

\begin{figure}[tbp]
 \begin{center}
   \mbox{\epsfig{figure=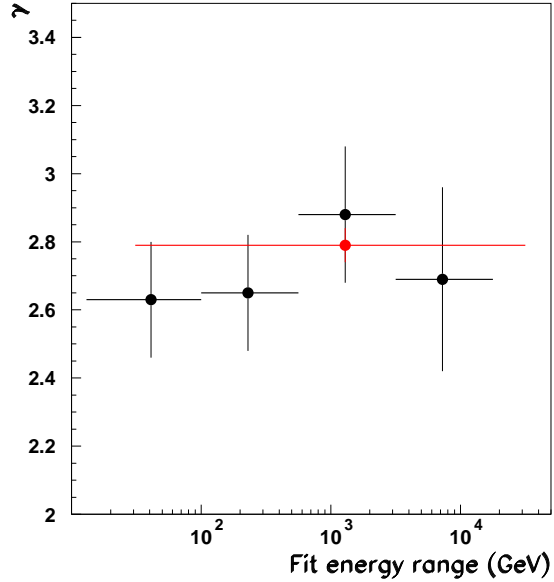,width=8.5cm,height=8.5cm}}
 \end{center}
\caption{\em {Slopes of the independent fits to the hadron flux.
The fitting energy range is shown. The circle and dashed line show
the slope as found in (\ref{eq:flufit}).}} \label{fi:slopes}
\end{figure}

A total systematic energy dependent uncertainty $\frac{\delta
S}{S} \simeq 15 \%$ is obtained from the first 3 items. To this,
the $15 \%$ constant systematic uncertainty due to item 4 has to
be
added.\\
The hadron flux is fitted by a power law from 30 GeV up to 30 TeV as \\
\begin{equation}
S_{h}(E_{h}) = (2.25 \pm 0.21) \times 10^{-7}
(\frac{E_{h}}{1000})^{(-2.79 \pm 0.05)} \ \ \ \
m^{-2}s^{-1}sr^{-1}GeV^{-1} \label{eq:flufit}
\end{equation}
with $\chi^2$ = 0.91 and is shown in Fig.\ref{fi:hadflu}. \\
In the fitting procedure (and in the plot), the energy dependent
systematic uncertainties have been included; the $15 \%$ energy
independent systematic effect
has to be added. \\
The hadron flux is compatible, within the errors, with a single
power law. This has been tested by performing  the same fit in
independent narrower energy ranges, the resulting slopes being
shown in Fig.\ref{fi:slopes}.

\section{\bf The primary proton spectrum}

\begin{figure}[tbp]
  \begin{center}
    \mbox{\epsfig{figure=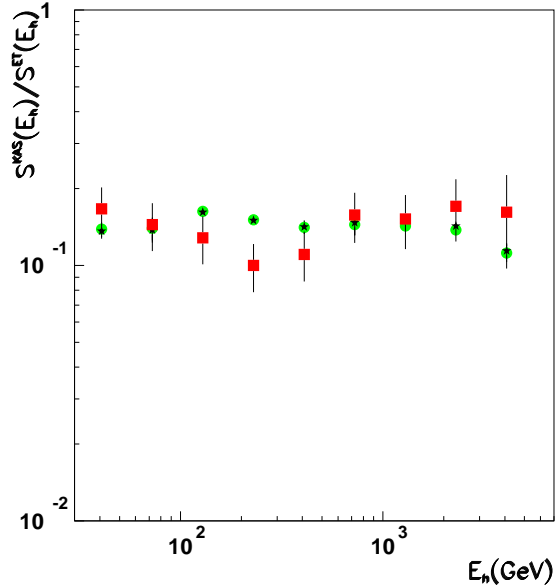,width=8.5cm,height=8.5cm}}
  \end{center}
\caption{\em{Ratio of the experimentally measured hadron fluxes by
EAS-TOP and KASCADE (squares) compared to the expectation if the
proton+helium primary spectra by JACEE (circles) or RUNJOB (stars)
are assumed.}} \label{fi:ratio}
\end{figure}

The primary proton spectrum is derived from the data by:\\
a) checking the hadron propagation code in the atmosphere; \\
b) subtracting from the measured hadron spectrum the contribution of heavy
primaries;\\
c) minimizing the difference between the measured and the
expected hadron fluxes on the basis of different primary proton spectra.

a) The region of interest coincides with the energy range in which
QGSJET (the hadronic interaction model used to describe the cosmic
ray interaction and propagation in the atmosphere) has been
directly checked against accelerator data \cite{engel,knapp}, both
concerning the leading particle and the secondary production
physics. Its reliability to reproduce the present data has been checked
by comparing its predictions to the measured ratio of hadron
fluxes at sea level (KASCADE \cite{casca}, 1030 g/cm$^2$) and
mountain altitude (EAS-TOP, 820 g/cm$^2$). Primary protons and
helium nuclei were generated in quasi vertical direction $\theta
\leq 5^{\circ}$, with energy spectra according to JACEE and RUNJOB
\cite{asa,apa} and the expected hadron fluxes at each observation
level were calculated. As shown in Fig.\ref{fi:ratio},
the expected ratio does not depend on the differences between such
primary spectra, and it is compatible with the measured one
within the statistical uncertainties, the comparison leading to
a $\chi^2$=1.2/d.o.f. On average, the model
reproduces the experimental ratio at better than 10$\%$.\\
We remind that the general features of the model relevant for the
calculation of the
hadron flux (and therefore object of the test) are the combination
of the total cross section and inelasticity concerning the contribution 
of the surviving primaries, and the very forward
production cross section for the contribution of the secondaries.
We therefore conclude that QGSJET, as implemented in CORSIKA, can
be reliably applied in the considered energy range in the
description of the uncorrelated hadron fluxes at different
atmospheric depths and therefore can be applied between the top of the
atmosphere and the EAS-TOP observation level, thus allowing to
derive the primary nucleon flux from the present measurement.

b) The contribution to the hadron flux from helium primaries has
been evaluated using their spectrum as directly measured by the balloon
experiments. In order to derive the systematic uncertainties of
the procedure, both the RUNJOB ($\gamma_{He}$ = 2.80) and JACEE
($\gamma_{He}$ = 2.68) data have been used and their contribution
subtracted from the experimental hadron flux. At $E_{h} \simeq$ 1
TeV, such contribution is $15 \%$ and $29 \%$ from RUNJOB and
JACEE respectively; the heavier nuclei one is less than 10 $\%$.

c)The primary proton spectrum is obtained as the one 
minimizing the difference between the measured hadron spectrum (after 
subtraction of the Helium contribution by means of the afore described 
procedure) and the
expected one from simulated proton primaries. Extensive
simulations have been carried on, generating primary protons in
quasi vertical direction ($\theta < 5^o$), with energy extracted
on power law spectra with slope varying between 2.5 and 3.2. The
number of simulated events is such that the number of hadrons in
each energy bin be much higher than the experimentally collected
one. 
Most of the contribution to each hadron energy bin comes from
different primary energy regions; hadrons of energies in different
ranges, e.g. $E_{h}$ = 0.1 $\div$ 0.2,  0.2 $\div$ 0.5, 0.5 $\div$
1, 1 $\div$ 2, 2 $\div$ 5, $\geq$ 5 TeV are produced by primaries
with median energy $E_{MED} \simeq$ 0.5, 2, 4, 10, 20, 55 TeV
respectively. The data thus allow to get information on the
primary proton spectrum in the range 0.5 $\div$ 50 TeV. \\
Assuming a primary spectrum of the power law form
$S(E_{0})=S_{0}E_{0}^{-\gamma}$, the
normalization factor $S_{0}$ and the slope $\gamma$ have
been obtained
minimizing the differences between the calculated and
the measured number of hadrons in each energy bin. 
The minimizations have been carried on by taking into
account both the statistical and the
energy dependent systematic uncertainties in the hadron flux. \\
\begin{figure}[tbp]
   \begin{center}
     \mbox{\epsfig{figure=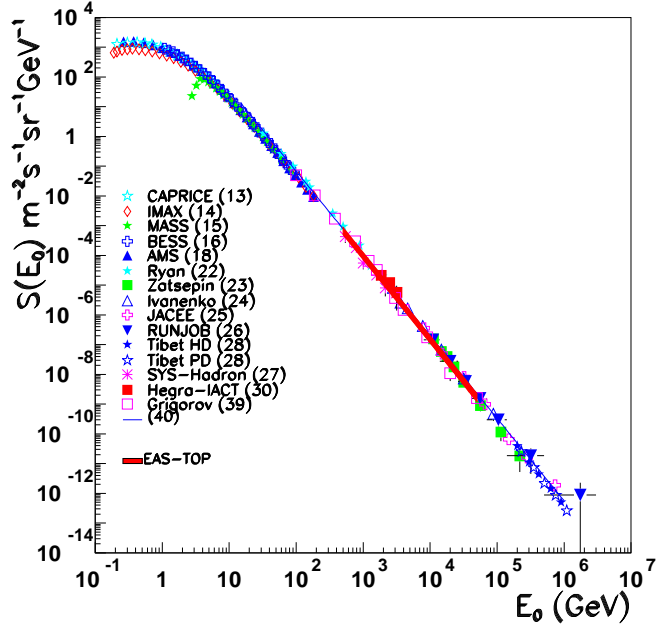,width=9.5cm,height=9.5cm}}
   \end{center}
\caption{\em{Primary proton spectrum; the full area represents the result
of this measurement and includes the systematic and statistical errors. 
Results from different experiments are shown for comparison. The straight 
line represents a fit from \cite{wbso}.}} 
\label{fi:prflu}
\end{figure}
\begin{figure}[tbp]
   \begin{center}
     \mbox{\epsfig{figure=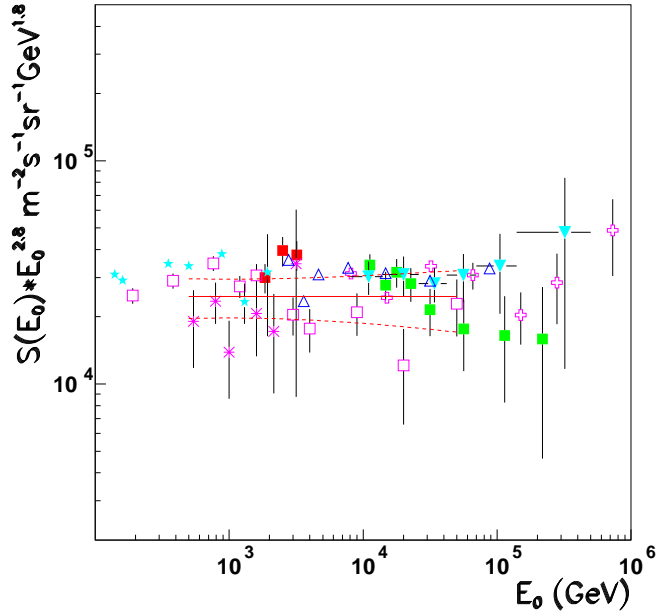,width=9.5cm,height=9.5cm}}
   \end{center}
\caption{\em{Primary proton spectrum multiplied by $E_{0}^{2.8}$.
Symbols as in Fig.\ref{fi:prflu}. The horizontal line represents the
present result, Eq.5, with errors.}} \label{fi:prflup}
\end{figure}
The data are well described by power law spectra in the energy
range 0.5 $\div$ 50 TeV, with best fits, for the case of subtraction of the
RUNJOB Helium spectrum:

$S(E_{0}) = (1.05 \pm 0.16) \times 10^{-4}(\frac{E_{0}}{1000})^{(-2.80 \pm 0.05)}$ 
m$^{-2}$ s$^{-1}$sr$^{-1}$ GeV$^{-1}$

and for the case of subtraction of the JACEE Helium spectrum:

$S(E_{0}) = (0.91 \pm 0.15) \times 10^{-4}(\frac{E_{0}}{1000})^{(-2.80 \pm 0.06)}$ 
m$^{-2}$ s$^{-1}$sr$^{-1}$ GeV$^{-1}$

Including the 7$\%$ uncertainty in the helium contribution and the
15$\%$ constant systematic uncertainty on the measured hadron flux
into a global systematic error term, the result can be summarized
as follows :

$S(E_{0}) = (9.8 \pm 1.1 \pm 1.6^{sys}) \times 10^{-5}(\frac{E_{0}}{1000})^{(-2.80 \pm 0.06)}$
m$^{-2}$s$^{-1}$sr$^{-1}$GeV$^{-1}$ \ \ \ (5)

The obtained proton spectrum is shown in Figs.\ref{fi:prflu} and
\ref{fi:prflup}. The full area and the shaded lines (in the two figures
respectively) include the systematic and statistical uncertainties of the 
measurement.

\section{\bf Conclusions}

The hadron flux has been measured over a wide energy range (30
GeV$\div$30 TeV) by means of the EAS-TOP hadron calorimeter at the
atmospheric depth of 820 g/cm$^{2}$. The spectrum is well
described by a single power law in the
whole range : \\
$S(E_{h}) = (2.25 \pm 0.21 \pm 0.34^{sys}) \times 10^{-7} (\frac{E
_{h}}{1000})^{(-2.79 \pm 0.05)}$ m$^{-2}$ s$^{-1}$ sr$^{-1}$ GeV$^{-1}$.\\
Taking into account the contamination from heavier nuclei, on the
basis of direct measurements, the primary proton spectrum is
obtained between 0.5 and 50 TeV and is found to be compatible with
a single
slope power law: \\
$S(E_{0}) = (9.8 \pm 1.1 \pm 1.6^{sys}) \times 10^{-5}
(\frac{E_{0}}{1000})^{(-2.80 \pm 0.06)}$
m$^{-2}$ s$^{-1}$ sr$^{-1}$ GeV$^{-1}$.\\
A systematic uncertainty of about 7 $\%$ due to the uncertainty in
the helium flux is included. The data match very well with the
direct measurements over a wide energy range, usually not
available to a single experiment,
where direct measurements become statistically poor.\\
The reliability of the CORSIKA/QGSJET interaction and propagation
code, which is used to propagate the hadrons in the atmosphere and
to compute the heavy nuclei contribution, is directly checked in
this energy range by comparison with accelerator data and,
concerning the direct application to the present measurement,
through its capability to reproduce the ratio of hadron fluxes as
measured at two different atmospheric depths by EAS-TOP and
KASCADE, at 820 and 1030 g/cm$^{2}$ respectively.

\section{Acknowledgements}
The cooperation of the Direction and Staff of the Gran Sasso
National Laboratories, as well as the technical assistance of
C.Barattia, R.Bertoni, G.Giuliani and G.Pirali are gratefully
acknowledged.\\
The comments of an unknown referee have contributed to improve and clarify
the text.

\end{document}